\documentclass[fleqn,usenatbib]{mnras}

\usepackage[T1]{fontenc}
\usepackage{ae,aecompl}
\usepackage{natbib,twoopt}
\usepackage{graphicx}	
\usepackage{amsmath}	
\usepackage{amssymb}	
\usepackage{amsfonts}
\usepackage{wasysym}
\usepackage{pdflscape}
\usepackage{comment}
\usepackage{float}
\usepackage{csquotes}
\usepackage{soul}

\title[Investigation of Rocket Effect in BRCs]{Investigation of Rocket Effect in Bright-Rimmed Clouds using \textit{Gaia} EDR3}

\author[Piyali Saha et al.]{
Piyali Saha,$^{1,2,3}$\thanks{E-mail: s.piyali16@gmail.com (PS)}
Maheswar G.,$^{2}$
D. K. Ojha,$^{4}$
Tapas Baug,$^{1}$
and Sharma Neha$^{5}$
\\
$^{1}$Satyendra Nath Bose National Centre for Basic Sciences, Salt Lake, Kolkata 700 106, India\\
$^{2}$Indian Institute of Astrophysics (IIA), Sarjapur Road, Koramangala, Bangalore 560034, India\\
$^{3}$Pt. Ravishankar Shukla University, Amanaka G.E. Road, Raipur, Chhatisgarh 492010, India\\
$^{4}$Tata Institute of Fundamental Research (TIFR), Homi Bhabha Road, Mumbai 400005, India\\
$^{5}$Aryabhatta Research Institute of Observational SciencES (ARIES), Manora Peak, Nainital 263002, India
}

\date{Accepted XXX. Received YYY; in original form ZZZ}

\pubyear{2022}

\begin{document}
\label{firstpage}
\pagerange{\pageref{firstpage}--\pageref{lastpage}}
\maketitle

\begin{abstract}
    Bright-rimmed clouds (BRCs) are excellent laboratories to explore the radiation-driven implosion mode of star formation because they show evidence of triggered star formation. In our previous study, BRC 18 has been found to accelerate away from the direction of the ionizing H{\sc ii} region because of the well known ``Rocket Effect". Based on the assumption that both BRC 18 and the candidate young stellar objects (YSOs) are kinematically coupled and using the latest \textit{Gaia} EDR3 measurements, we found that the relative proper motions of the candidate YSOs exhibit a tendency of moving away from the ionizing source. Using BRC 18 as a prototype, we made our further analysis for 21 more BRCs, a majority of which showed a similar trend. For most of the BRCs, the median angle of the relative proper motion of the candidate YSOs is similar to the angle of on-sky direction from the ionizing source to the central \textit{IRAS} source of the BRC. Based on Pearson's and Spearman's correlation coefficients, we found a strong correlation between these two angles, which is further supported by the Kolmogorov-Smirnov (K$-$S) test on them. The strong correlation between these two angles supports the ``Rocket Effect'' in the BRCs on the plane-of-sky. 
\end{abstract}




\section{Introduction}

The origination and evolution of massive stars can immensely affect the immediate surroundings where they form. The strong stellar winds or some supernova explosions may compress the nearby molecular clouds and thus trigger a subsequent formation of stars that otherwise may not have initiated. This is typically known as positive stellar feedback. In a converse scenario, the effect may be too intense that results in hindering further star formation activity by dispersing the pre-existing cloud material, which can be considered as negative stellar feedback. Thus stellar feedback has significant aftermaths in terms of cloud energetics, multiple stellar population, regulation of star formation rates, star formation efficiency, and so on \citep{2007ARA&A..45..481Z}.

The strong ultraviolet radiation emitting from the massive stars, or expansion of H{\sc ii} regions, develops high-pressure waves by the ionizing heating into the surrounding pre-existing dense clumps. This enhanced pressure may trigger the birth of new proto-stellar cores or compress the pre-existing ones to eventually lead to formation of a new generation of stars. This mechanism of star formation triggered by radiation from neighbouring massive stars is known as radiation-driven implosion \citep[RDI;][]{1989ApJ...346..735B, 1990ApJ...354..529B}. The consequent clouds are known as bright-rimmed clouds \citep[BRCs; ][]{1991ApJS...77...59S, 1994ApJS...92..163S}, having their illuminated rim facing towards the ionizing star(s) and often tied up with an elongated structure along the opposite direction. They are isolated molecular clouds and often found to be located at the boundaries of evolved H{\sc ii} regions. \cite{1991ApJS...77...59S} and \cite{1994ApJS...92..163S} listed a total of 89 BRCs based on their association with the Infrared Astronomical Satellite (\textit{IRAS}) sources. Observational evidence of star formation, e.g., \textit{IRAS} point sources, H$\alpha$ emitting stars, infrared (IR) excess sources, Herbig-Haro objects, etc., were reported in those BRCs \citep[e.g., Table 1 of][]{1998ASPC..148..150E}. Also, signposts of spatial distribution along with gradient in evolutionary stages of young stellar objects (YSOs) are detected along the axes of the BRCs, which is typically known as ``small-scale sequential star formation" \citep[e.g., ][]{1995ApJ...455L..39S, 2007PASJ...59..199O, 2010ApJ...717.1067C}. Thus, BRCs are considered to be the markers of positive stellar feedback in the formation of YSOs.

Several analytical studies were carried out to understand the evolution of clouds during the RDI mode \citep[e.g., ][]{1955ApJ...121....6O, 1969Phy....41..172K}. The process starts when an ionization front preceded by a shock front travels across the dense globule, causing it to collapse and resulting in formation of a highly dense core, which eventually could trigger the star formation. The time at which the star formation begins and the fraction of cloud material gets transformed into stars is typically $\sim10^{5}$ yrs \citep[][hereafter, Paper I]{2022MNRAS.510.2644S}. At the surface of ionization front, because of the enhanced pressure the ionized cloud material starts photo-evaporating towards the ionizing source and as a back reaction of the expelled ionized gas, the residual cloud escalates away from the direction of ionizing H{\sc ii} region, which is typically known as the ``Rocket Effect'' \citep[hereafter, RE; ][]{1954BAN....12..177O}. The YSOs formed by consequence of the RDI should share the kinematics similar to the accelerating cloud. Thus, the YSOs would accelerate radially away from the nearby massive star/s, which is/are responsible for ionizing the cloud and triggering the star formation in it \citep{2015MNRAS.450.1199D}.

\begin{figure}
	\scriptsize
	\centering
	\includegraphics[height=4.5cm, width=7cm]{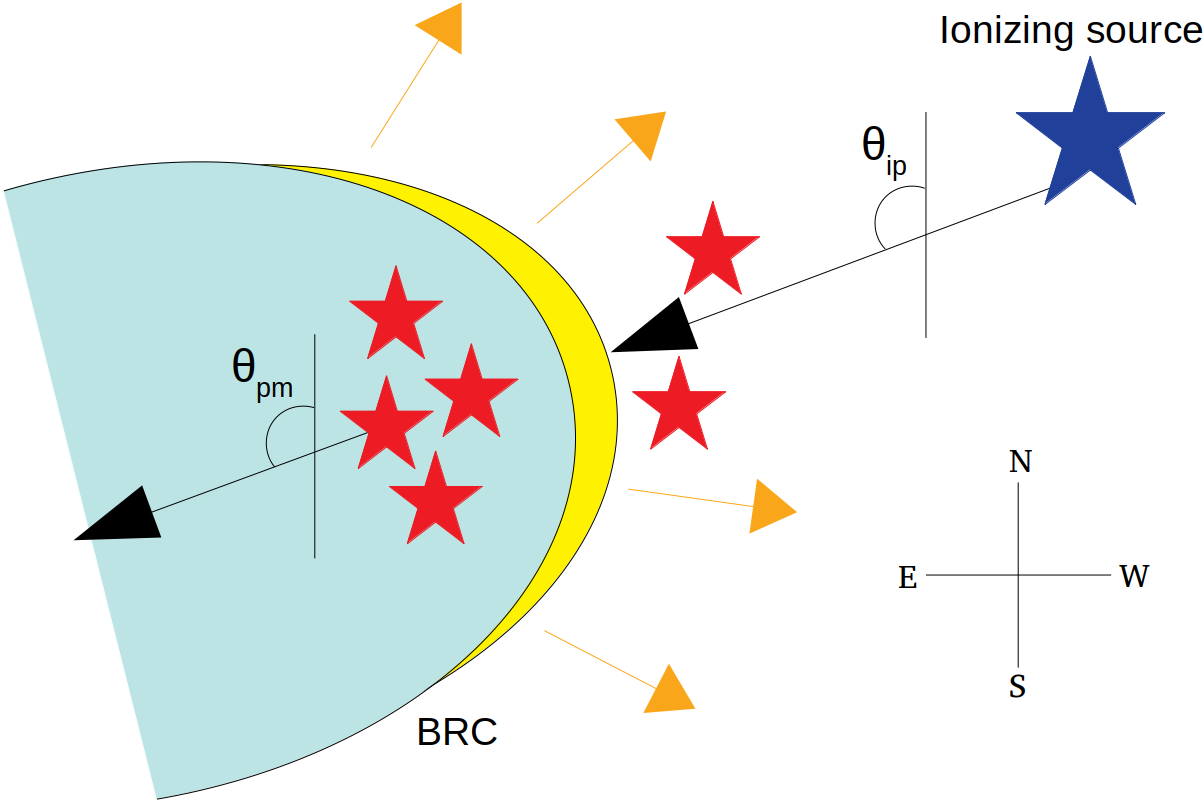}
	\caption{An illustration of a system consisting of a BRC (blue cloud with a thick yellow colored rim) and an ionizing source (blue star) (not to scale). The YSOs distributed towards the BRC are presented using red star symbols. The direction of photo-evaporation flow of the cloud material is shown using yellow colored arrows. The black colored arrow making an angle $\theta_{ip}$ with respect to the north and directing away from the ionizing star, is the direction of ionizing photons. Another black colored arrow making an angle $\theta_{pm}$ with respect to the same and directing away from the YSOs, presents the angle of their relative proper motion.}
	\label{fig:model_rdi}
\end{figure}

To investigate the RE on the BRCs, we first examined the proper motion (PM) of candidate YSOs associated with BRC 18 in Paper I. BRC 18 resides at the eastern periphery of the $\lambda$ Ori H{\sc ii} region. The strong ionizing photons coming from the earliest type star of the Collinder 69 cluster, $\lambda$ Ori \citep[O8III; ][]{1974ApJ...193..113C}, is considered to be responsible for the ongoing RDI mechanism in BRC 18 \citep{1991ApJS...77...59S}. Using the latest \textit{Gaia} Early Data Release 3 (EDR3) astrometric measurements, we found that the majority of the candidate YSOs located towards BRC 18 share similar PM values and are lying at similar distances suggesting that they are all formed together and currently moving as a group. We then subtracted the median PM of the $\lambda$ Ori H{\sc ii} region from the observed PMs of the candidate YSOs vectorially. This results in the true internal projected motions of these candidate YSOs. A similar approach was followed in other recent studies also, for example, \citet{2019A&A...622A.118S} and \citet{2021MNRAS.507..267A}. Fig. \ref{fig:model_rdi} shows a cartoon diagram of a system consisting of a BRC and one ionizing source (also see Paper I), where we depict the angles used in our analysis. $\theta_{ip}$ and $\theta_{pm}$ represent the angle of ionizing photons and the angle of the internal projected motion of YSOs, respectively. The $\theta_{pm}$ is estimated by the median angle made by the relative PM or internal projected motion of the candidate YSOs, starting from celestial north and increasing eastward. We estimate the $\theta_{ip}$ as the angle made by the line joining the ionizing source and the central \textit{IRAS} source residing in each BRC (Paper I). The $\theta_{ip}$ values are also measured following the same convention as it was done for $\theta_{pm}$. As discussed in Paper I, in an ideal RE scenario, $\theta_{ip}$ is expected to be similar to $\theta_{pm}$, i.e., $\mid$$\theta_{ip}-\theta_{pm}$$\mid\simeq0\degr$, on the sky plane.

Both BRC 18 and Collinder 69 cluster, of which $\lambda$ Ori is a part, reside at a similar distance ($\sim400$ pc; Paper I) indicating that both are lying almost on the sky plane. In BRC 18, we found $\mid$$\theta_{ip}-\theta_{pm}$$\mid=1\degr\pm14\degr$, which is very close to 0$\degr$, supporting the RE. Motivated by this finding we investigated the RE in additional 21 BRCs having a considerable number of candidate YSOs with reliable counterparts in the \textit{Gaia} EDR3 archive. This paper is structured in the following manner. The archival data and data from literature are explained in section \ref{sec:sel_brcs}. A relevant discussion on the results is presented in section \ref{sec:dis}. Finally, we summarize and conclude our work in section \ref{sec:con}.

\vspace{-0.2cm}
\section{Archival Data and data from literature}\label{sec:sel_brcs}

Based on various methods, for example, Two Micron All Sky Survey (2MASS), \textit{Spitzer}, \textit{WISE} archival data and results from near- and mid-IR photometric observations \citep[e.g.,][]{2012PASJ...64...96H, 2014MNRAS.443.1614P, 2016AJ....151..126S}, slitless spectroscopy \citep{2002AJ....123.2597O, 2008AJ....135.2323I, 2020arXiv200600219H}, optical BVI$_{c}$ photometry \citep[e.g.][]{2009MNRAS.396..964C}, optical spectroscopy \citep[e.g.][]{2010ApJ...717.1067C, 2018AJ....156...84K}, X-ray imaging \citep{2012MNRAS.426.2917G}, astrometry from \textit{Gaia} data release 2 (DR2) \citep{2018AJ....156...84K, 2018A&A...620A.172Z}, method of maximum likelihood \citep{2018MNRAS.477..298G}, presence of the candidate YSOs in the BRCs has been reported. We collected information of all candidate YSOs available in these references for our study.

\textit{Gaia} EDR3 \citep{2020arXiv201201533G} presents precisely measured astrometric parameters of more than one billion objects, brighter than $G\sim21$. We acquired distances (\textit{d}) and PMs in right ascension ($\mu_{\alpha\star}$) and declination ($\mu_{\delta}$) from \cite{2021AJ....161..147B} and the \textit{Gaia} EDR3 archive \citep{2020arXiv201201533G}, respectively, for our sources using a cross-match radius of 1$\arcsec$. Sources having their ratios, $d/\Delta d$, $\mu_{\alpha\star}/\Delta\mu_{\alpha\star}$, and $\mu_{\delta}/\Delta\mu_{\delta}$ all are $\geqslant3$, are considered for this study, where, $\Delta d$, $\Delta\mu_{\alpha\star}$, and $\Delta\mu_{\delta}$ are the measurement uncertainties in $d$, $\mu_{\alpha\star}$, and $\mu_{\delta}$, respectively. In an unresolved multiple stellar system, the presence of an orbiting companion around a source introduces a wobble at the point of maximum flux. Thus, the barycenter of the stellar system shifts away from its photocenter. This, in turn, results in a poorer goodness-of-fit statistics, and comes up with a higher renormalized unit weight error (RUWE), which is considered as a reliable indicator of the existence of a nearby associate \citep[e.g., ][]{2020MNRAS.496.1922B, 2021ApJ...907L..33S, 2022A&A...657A...7K}. We included only those sources for which the RUWE $\leq$1.4 \citep{LL:LL-124}. The selection criteria are also mentioned in Paper I.

\section{Results and Discussions}\label{sec:dis}

\subsection{Selection of the BRCs}\label{subsec:sel_brcs}

The methodology adopted here to affirm the coupling between young sources with their parental clouds is similar to our previous studies \citep{2020MNRAS.494.5851S, 2021A&A...653A.142S, 2022MNRAS.510.2644S}. Based on the literature, out of 89 BRCs, candidate YSOs have been identified towards 52 of them using various methods till date (as described in section \ref{sec:sel_brcs}). Among the 52 BRCs, we selected 29 of them, which are associated with at least 4 candidate YSOs satisfying our selection criteria. The median absolute deviation (MAD) was used to compute the statistical dispersion in $d$, $\mu_{\alpha\star}$, and $\mu_{\delta}$. Only those candidate YSOs, which are lying within 5$\times$MAD with respect to the median values of the $d$, $\mu_{\alpha\star}$, and $\mu_{\delta}$, are selected as the members of each of their respective BRCs. The adoption of 5$\times$MAD was made because the majority of the candidate YSOs in the BRCs are distributed within this boundary. We recomputed the median and the MAD of $d$, $\mu_{\alpha\star}$, and $\mu_{\delta}$ of the selected candidate YSOs to estimate the final astrometric results of the BRCs (also see Paper I). 

Of the 29 BRCs, we first excluded BRCs 15, 24, and 30, where the candidate YSOs are distributed far from the cloud-ionizing star vicinities. Also, we excluded BRC 51 due to its high MAD in $\mu_{\delta}$ (-2.187$\pm$6.657 mas yr$^{-1}$, obtained from \textit{Gaia} EDR3 measurements of the candidate YSOs) from further study. Also, ionizing stars of this BRC \citep[$\zeta$ Pup and $\gamma^{2}$ Vel; ][]{2004A&A...415..627T} do not have astrometric measurements in \textit{Gaia} EDR3. Additionally, we set a numerical limit $z=(d-d_{\star})/(\Delta d+\Delta d_{\star})$ for the consistency of distance between the BRC (or the candidate YSOs residing in the BRC) and the respective ionizing sources, where, $d_{\star}$ and $\Delta d_{\star}$ are the distance for the ionizing source and its corresponding uncertainty, respectively. In our study, we select only those BRCs, for which $z\leq1$. Thus, BRCs 34, 68, and 76 were eliminated. Therefore, out of 29 BRCs, seven were excluded from our further analysis. In Tables 1 and 2 (available as supplementary material), we present the details of candidate YSOs in the BRCs along with the $\theta_{pm}$ and $\theta_{ip}$ found in the analysis. The strategies followed to obtain the $\theta_{pm}$ of the candidate YSOs in the remaining 22 BRCs are discussed below:

\begin{figure*}
	\includegraphics[height=4.4cm, width=5.8cm]{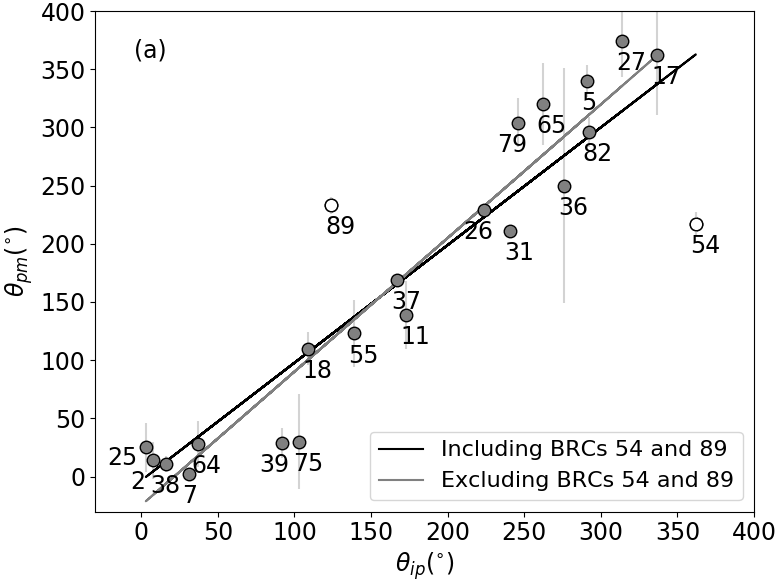} 
	\includegraphics[height=4.4cm, width=5.7cm]{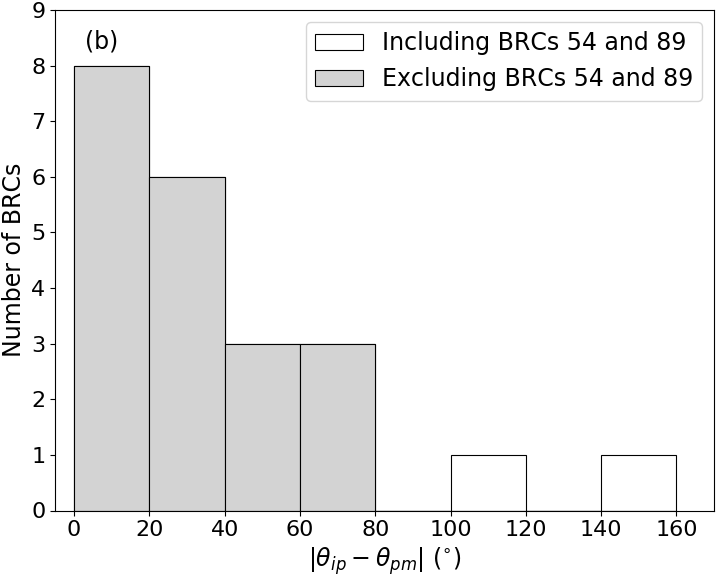}
	\includegraphics[height=4.4cm, width=5.7cm]{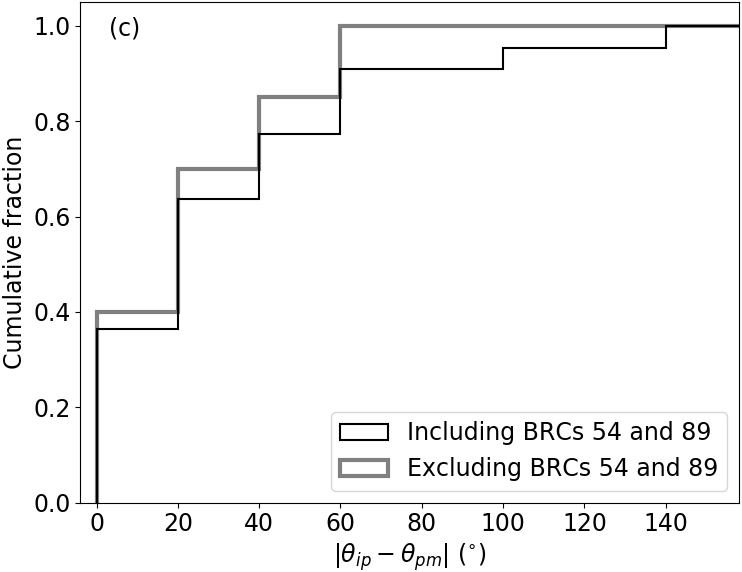} 
	\caption{\textbf{(a)} Plot of $\theta_{ip}$ versus $\theta_{pm}$. 
	The solid black line shows the best-fitted line for 22 BRCs (case A). The solid grey line shows the best-fitted line for 20 BRCs, excluding BRCs 54 and 89 (case B). \textbf{(b)} Histogram of $\mid$$\theta_{ip}-\theta_{pm}$$\mid$ for the BRCs with bin size $\sim20\degr$. \textbf{(c)} Cumulative histogram of the distribution of $\mid$$\theta_{ip}-\theta_{pm}$$\mid$ for the BRCs with bin size $\sim20\degr$.}
	\label{fig:plot_angle}
\end{figure*}

\textbf{BRC 2:} The massive source responsible for ionization of BRC 2 is BD+66 1675, which does not fulfill our selection criteria. We used the median PM obtained from the members of the H{\sc ii} region Sh2-171, which are associated with BD+66 1675 \citep{2018MNRAS.477..298G, 2018AJ....155...44P}, as the reference to obtain the $\theta_{pm}$.

\textbf{BRCs 5 and 7:} Three early type sources, BD+60 502, BD+60 504, and BD+60 507, are considered to influence BRCs 5 and 7 \citep{1970PASJ...22..277I, 2004A&A...426..535M}. BD+60 504, and BD+60 507 both have good astrometric measurements and reside at 2122$_{-81}^{121}$ and 1960$_{-68}^{65}$ pc, respectively. We used the PMs of BD+60 504 as the reference as its $d$ is more consistent with the BRCs. 

\textbf{BRCs 17 and 18:} BRCs 17 and 18 reside at the $\lambda$ Ori H{\sc ii} region and considered to be ionized by $\lambda$-Ori \citep{1991ApJS...77...59S}. The methodology adopted to find the relative PM of the candidate YSOs in these BRCs is described in Paper I. 

\textbf{BRC 25:} The massive source considered to ionize BRC 25 is HD 47839, which has a higher RUWE making its astrometric measurements unreliable. Therefore, we obtained the reference PM values of NGC 2264 from \cite{2020A&A...636A..80B} to estimate the $\theta_{pm}$.

\textbf{BRCs 36 to 39:} BRCs 36-39, located in the H{\sc ii} region IC 1396, are considered to be ionized by the central ionizing source HD 206267 \citep{1991ApJS...77...59S}, which fails our selection criteria. Therefore, we used median PM values of the cluster members of IC 1396 provided in the literature \citep{2002AJ....124.1585C, 2005AJ....130..188S, 2009AJ....138....7M,2019A&A...622A.118S} as the reference to compute the $\theta_{pm}$ of the candidate YSOs in the BRCs. 

\textbf{BRC 55:} For BRC 55, in the H{\sc ii} region RCW 27, the massive source responsible for ionizing BRC 55 is HD 73882, which has RUWE = 2.068. Therefore, we obtained the astrometric parameters of the members of RCW 27 \citep{2018A&A...617A..63P} from \textit{Gaia} EDR3 and used the median PM of the cluster members as the reference PM to estimate $\theta_{pm}$. 

\textbf{BRC 64:} The star responsible for ionizing BRC 64 is LSS2231, located at 3145$_{-437}^{400}$ pc, while BRC 64 resides at 2968$\pm$225 pc. Thus the lower limit of $d$ of LSS2231 lies within the distance range of BRC 64. Though the RUWE of LSS2231 is 1.819, this is to note that it satisfies all of our other selection criteria. Due to lack of information of any other members of the H{\sc ii} region BBW 347 where BRC 64 resides, we used the PM values of LSS2231 to estimate the $\theta_{pm}$.

\textbf{BRC 65:} There are three early type sources considered to ionize BRC 65, which are, HD 101131, HD 101205, and HD 101436 \citep{2004A&A...415..627T}, located in the H{\sc ii} region RCW 62. As only HD 101205 satisfies our selection criteria, we used its PM values as the reference to compute the $\theta_{pm}$. 

\textbf{BRC 79:} BRC 79 is found to be located in the H{\sc ii} region RCW 108. Two massive sources are found to be responsible for the ionization of this region, HD 150136 (O5III) and HD 150135 (O6.5V) \citep{2004A&A...415..627T}. As $d$ of HD 150136 (1428$_{-96}^{93}$ pc) is more consistent than HD 150135 (1242$_{-39}^{37}$ pc) with the same for BRC 79, and the spectral type of HD 150136 is earlier than the later, we considered the PM of HD 150136 as the reference PM. 

\textbf{BRC 82:} Three early type sources, HD 152233 (O6III), HD 326286 (B0), and HD 152245 (B0Ib) are considered to be responsible for the ionization of BRC 82 \citep{1999PASJ...51..791Y,2004A&A...415..627T}. The earliest type ionizing source, HD 152233, is a member of the cluster NGC 6231. As there is a high possibility of the influence of other massive stars of NGC 6231, we obtained median PM of the cluster members from \cite{2017AJ....154...87K} and considered those as reference PM. 

The rest BRCs, i.e., 11, 26, 27, 31, 54, 75, and 89 are associated with H{\sc ii} regions having reliable astrometric measurements of the ionizing sources in the \textit{Gaia} EDR3 catalogue. Therefore, the PMs of these ionizing sources were used as the reference PMs in order to investigate the RE taking place in the respective BRCs.

\subsection{Internal projected motion of the candidate YSOs in the BRCs}\label{subsec:rel_pm}

Our study finally focuses on 22 BRCs (including BRC 18), which consist of a considerable number of candidate YSOs with reliable astrometric measurements in \textit{Gaia} EDR3. Mostly the YSOs are embedded deep in the natal cloud, so the optical emission from many of them are highly obscured by the circumstellar and interstelllar dust resulting in unreliable or no detection in \textit{Gaia} EDR3. In Fig. \ref{fig:plot_angle} (a) we present the distribution of $\theta_{ip}$ versus $\theta_{pm}$ of 22 BRCs. BRCs 54 and 89 show distinctively higher $\mid$$\theta_{ip}-\theta_{pm}$$\mid$ compared to other BRCs. We marked them using open circles. Other 20 BRCs are shown using filled grey circles. We made two cases A and B, which indicate inclusion and exclusion of both BRCs 54 and 89, respectively. The solid black line shows the best-fitted line for case A. The slope of this line is estimated as 1.011$\pm$0.103. We again show the best-fitted line in case B using solid grey line, of which the slope is estimated as 1.148$\pm$0.071. Error in $\theta_{pm}$ is estimated using the error propagation method. The error in $\theta_{ip}$ is negligible because the positional errors of the sources are in milli arcsecond order. To obtain the correlation between $\theta_{ip}$ and $\theta_{pm}$, we computed the Pearson's correlation coefficient. For case A, the value is found to be 0.909 with pnull 4.641$\times$10$^{-9}$. The Spearman's correlation coefficient is estimated as 0.899, with pnull as 1.247$\times$10$^{-8}$. We also performed Kolmogorov-Smirnov (K$-$S) test on these angles. The computed statistic is 0.182, and the pvalue is 0.821. For case B, the Pearson's and Spearman's correlation coefficients for 20 BRCs are estimated as 0.967 and 0.964 with pnull 3.444$\times$10$^{-12}$ and 8.654$\times$10$^{-12}$, respectively. The K$-$S test provides statistic of 0.200, and the pvalue is 0.771 for the 20 BRCs. All these statistical analyses indicate a strong correlation between $\theta_{ip}$ and $\theta_{pm}$. The distribution of $\mid$$\theta_{ip}-\theta_{pm}$$\mid$ as a form of histogram with binsize $20\degr$ is shown in Fig. \ref{fig:plot_angle} (b), using white- and grey-colored bins, for cases A and B, respectively. A maximum in the lowest of $\mid$$\theta_{ip}-\theta_{pm}$$\mid$ and a decrease in the number of BRCs with higher $\mid$$\theta_{ip}-\theta_{pm}$$\mid$ is clearly noticeable. In Fig. \ref{fig:plot_angle} (c), we show the cumulative histogram of $\mid$$\theta_{ip}-\theta_{pm}$$\mid$, using white- and grey-colored bins, for cases A and B, respectively. It is apparent from both Figs. \ref{fig:plot_angle} (b) and (c), that almost $\sim80$\% of the BRCs show $\mid$$\theta_{ip}-\theta_{pm}$$\mid\leq60\degr$. 

This is to note that because of the lack of information of the radial velocities of the sources, in Paper I as well as in this study too, we consider only the PMs of the sources, which is on the sky plane. In an H{\sc ii} region, where the ionizing star and the BRC both are not located at the sky plane, but have a significant distance along the line-of-sight, the $\mid$$\theta_{ip}-\theta_{pm}$$\mid\sim0\degr$ might not always be satisfied because of the possible presence of a significant radial component of motion of the BRC/ionizing source. Apparently, $180\degr>\mid$$\theta_{ip}-\theta_{pm}$$\mid>90\degr$ might seem as a phenomenon opposite to RE in 2D, i.e. PMs of the YSOs towards the ionizing star, but these sources may have significant components of motions in the radial direction, which might make their total motions in 3D away from the ionizing star, satisfying the RE scenario. Thus, the $\mid$$\theta_{ip}-\theta_{pm}$$\mid\sim0\degr$ condition does not signify the RE scenario completely in 2D, but in 3D.

A higher $\mid$$\theta_{ip}-\theta_{pm}$$\mid$ for BRCs 54 and 89 signifies that the relative PM of the candidate YSOs direct towards their respective ionizing sources. For BRC 54, we initially suspected that the presence of another massive star, HD 73882 (also considered to ionize BRC 55), might have influenced the motion of the candidate YSOs. But due to higher distance ($\sim$13 pc), the probability of HD 73882 influencing BRC 54 is low. \cite{2018A&A...617A..63P} detected second generation of stars around BRC 54, which indicates that the kinematic coupling between the candidate YSOs and BRC 54 might have been lost. Therefore, the $\theta_{pm}$ of the candidate YSOs might not indicate the PM of BRC 54. 

The massive star considered to ionize BRC 89 is HD 165921, with respect to which $\theta_{pm}$ of the candidate YSOs is estimated. Based on the 2MASS all-sky survey, \cite{2003A&A...404..223B} discovered a stellar cluster encircling an ionizing source HD 166056, a B2Ve star \citep{1982AJ.....87...98H} in BRC 89. It is embedded in a cavity, surrounded by a prominent shell-like structure indicative of expanding ionization front \citep{2014ApJ...783....1S}. The pressure of HD 166056 might have affected the motion of the surrounding YSOs which is why $\theta_{pm}$ of the candidate YSOs show significantly different direction in BRC 89.

\section{Summary and Conclusions}\label{sec:con}

As an effect of the RDI mechanism, the compression of the BRCs because of ionization by the nearby OB stars results into triggered star formation. This leads to a group of young population staying behind the BRCs. Also, the BRCs escalate away from the direction of ionizing H{\sc ii} region as consequence of the photo-evaporation flow of cloud material in the direction of H{\sc ii} region, which is well-known as the Rocket Effect. Assuming that the internal projected motion of the YSOs imitates the same of their parental clouds in the sky plane, 22 BRCs (including BRC 18; Paper I) are investigated based on the astrometric measurements of the associated candidate YSOs obtained from the latest \textit{Gaia} EDR3. As BRCs 54 and 89 show relatively higher $\mid$$\theta_{ip}-\theta_{pm}$$\mid$, we made two cases A and B, which indicates inclusion and exclusion of both the BRCs, respectively. The main outcomes of this work are summarized below:
\begin{itemize}
	\item The Pearson's correlation coefficient of $\theta_{ip}$ and $\theta_{pm}$ for case A is found to be 0.909 with pnull 4.641$\times$10$^{-9}$. For case B, the same is estimated as 0.967 with pnull 3.444$\times$10$^{-12}$.
	\item The Spearman's correlation coefficient is found to be 0.899, with pnull as 1.247$\times$10$^{-8}$ for case A. For case B, the same is found to be 0.964 with pnull 8.654$\times$10$^{-12}$.
	\item The statistic of K$-$S test is 0.182 with pvalue 0.821 and 0.200 with pvalue 0.771, for cases A and B, respectively.
	\item Based on the histogram of $\mid$$\theta_{ip}-\theta_{pm}$$\mid$, we found that most of the BRCs have a minimum of the difference between $\theta_{ip}$ and $\theta_{pm}$. 
	\item The strong correlation between $\theta_{ip}$ and $\theta_{pm}$ supports the RE in most of the BRCs on the plane-of-sky.
\end{itemize}
As the operating wavelength of \textit{Gaia} is optical, the number of candidate YSOs detected is less possibly due to the presence of dense gas and dust surrounding them. More deep observations with good quality data would help us to increase the statistics and a better understanding.

\section*{Acknowledgements}

We are grateful to the referee, Dr. Alexander Slater Binks, for his careful examination and generous suggestions that considerably improved the quality of the manuscript. We acknowledge the support by the S. N. Bose National Centre for Basic Sciences under the Department of Science and Technology, Govt. of India. This work has made use of archival data from the European Space Agency (ESA) mission \textit {Gaia} (\url{https://www.cosmos.esa.int/gaia}), processed by the \textit{Gaia} Data Processing and Analysis Consortium (DPAC, \url{https://www.cosmos.esa.int/web/gaia/dpac/consortium}). Funding for the DPAC has been provided by national institutions, in particular the institutions participating in the {\it Gaia} Multilateral Agreement (MLA). DKO acknowledges the support of the Department of Atomic Energy, Government of India, under Project Identification No. RTI 4002.

\section*{Data Availability}

The \textit{Gaia} EDR3 proper motions and distances of the sources are available in \url{https://vizier.u-strasbg.fr/viz-bin/VizieR-3?-source=I/350/gaiaedr3} and \url{https://vizier.u-strasbg.fr/viz-bin/VizieR?-source=I/352}, respectively. 



\bibliographystyle{mnras}
\bibliography{reference} 


\begin{table*}
	\begin{center}
	\caption{Kinematic properties of the candidate YSOs in the BRCs and the respective ionizing sources.}
	\label{tab:angles_list_gaia_brc}
	\renewcommand{\arraystretch}{1.3}
	\footnotesize
	\begin{tabular}{llllll}
		\hline
		 BRC Name$^{*}$ &  H{\sc ii} region & Ionizing Star & $d$ (pc)& $\theta_{\rm pm} (^{\circ})$ & $\theta_{\rm ip}(^{\circ})^{\dagger}$\\
		(1)&(2)&(3)&(4)&(5)&(6)\\
        \hline
        SFO 2 & Sh2-171 & BD+66 1675 & 1027$\pm$69 & 14$\pm$16&8 \\
        SFO 5 & IC 1805 & BD+60 504 & 2007$\pm$236 & 340$\pm$14&291\\
        SFO 7 & IC 1805 & BD+60 504 & 2192$\pm$31 & 2$\pm$4&31\\
        SFO 11 & IC 1848 & HD 17505 & 2119$\pm$237 & 139$\pm$29&173 \\
        SFO 17 & Sh2-264 & $\lambda$ Ori & 389$\pm$9 & 362$\pm$51&337 \\
        SFO 18$^{\dagger}$ & Sh2-264 & $\lambda$ Ori & 394$\pm$7 & 110$\pm$14&109 \\
        SFO 25 & NGC 2264 & HD 47839 & 692$\pm$12 & 25$\pm$21&3 \\
        SFO 26 & Sh2-296 & HD 54662 & 1236$\pm$127 & 229$\pm$3&224\\
        SFO 27 & Sh2-296 & HD 53456 & 1095$\pm$71 & 374$\pm$31&314\\
        SFO 31 & Sh2-117 & HD 199579 & 787$\pm$51 & 211$\pm$4&241 \\
        SFO 36 & Sh2-131 & HD 206267 & 902$\pm$47 & 250$\pm$101&276 \\
        SFO 37 & Sh2-131 & HD 206267 & 968$\pm$37 & 169$\pm$5&167 \\
        SFO 38 & Sh2-131 & HD 206267 & 923$\pm$41 & 11$\pm$7&16\\
        SFO 39 & Sh2-131 & HD 206267 & 932$\pm$47 & 29$\pm$13&92\\
        SFO 54 & NGC 2626 & vdBH17a & 978$\pm$99 & 217$\pm$10&362 \\
        SFO 55 & RCW 27 & HD 73882 & 897$\pm$32 & 123$\pm$29&139\\
        SFO 64 & BBW 347 & LSS2231 & 2968$\pm$225 & 28$\pm$20&37 \\
        SFO 65 & RCW 62 & HD 101205 & 2434$\pm$78 & 320$\pm$35&262 \\
        SFO 75 & RCW 98 & LSS 3423 & 1884$\pm$422& 30$\pm$41&103\\
        SFO 79 & RCW 108 & HD 150136 & 1515$\pm$263 & 304$\pm$21&246 \\
        SFO 82 & RCW 113/116 & HD 152233 & 1804$\pm$258& 296$\pm$13&292 \\
        SFO 89 & Sh2-29 & HD 165921 & 1290$\pm$1394& 233$\pm$7&124 \\
        \hline
	\end{tabular}\\
	\end{center}
    \begin{flushleft}
    \footnotesize {$^{*}$ BRCs are identified as SFO (Sugitani, Fukui, Ogura) in Simbad.\\
	$^{\dagger}$ Error in $\theta_{\rm ip}$ is negligible, 
	hence not tabulated.\\
	BRCs having $\theta_{\rm ip}$ and $\theta_{\rm pm}>360^{\circ}$ are actually corrected for keeping minimum angular distance between these two. For example, BRC 17 has $\theta_{\rm pm}=2^{\circ}\pm51^{\circ}$, but as $\theta_{\rm ip}=337^{\circ}$, we added $360^{\circ}$ with $\theta_{\rm pm}$ to estimate the minimum $\mid$$\theta_{ip}-\theta_{pm}$$\mid$.}\\
    \end{flushleft}
\end{table*}
\begin{table}
	\centering
	\caption{Kinematic properties of the candidate YSOs in the BRCs and the respective ionizing sources.}
	\label{tab:sources_list_gaia_brc}
	\footnotesize
	\begin{tabular}{ccc}
		\hline
		 candidate YSOs (\textit{Gaia} EDR3) & RA & Dec.\\
		(1)&(2)&(3)\\
        \hline
        \multicolumn{3}{c}{SFO 2}\\
        529108978383373824 & 0.817180 & 68.533421\\
        529108913962819840 & 0.822722 & 68.528244\\
        529116984202552064 & 0.891505 & 68.665579\\
        529109562499260672 & 0.996169 & 68.546432\\
        529109669877060096 & 0.999364 & 68.561558\\
        529109704236796288 & 1.006998 & 68.570463\\
        529110082193916416 & 1.007424 & 68.576971\\
        529109704236795136 & 1.010965 & 68.573980\\
        529110009175851648 & 1.019057 & 68.581070\\
        529109631218432896 & 1.021873 & 68.564734\\
        529109635517322240 & 1.023521 & 68.562283\\
        529109631218428160 & 1.030511 & 68.561745\\
        529109596858989184 & 1.031638 & 68.556941\\
        529107981950942080 & 1.041955 & 68.514546\\
        529109631218727680 & 1.048587 & 68.557023\\
        529109596859011200 & 1.051103 & 68.546706\\
        529109493779803392 & 1.058296 & 68.539329\\
        529109394995715584 & 1.061381 & 68.546951\\
        529109291919950080 & 1.062938 & 68.538589\\
        529107226036641024 & 1.066895 & 68.457339\\
        529107638353526912 & 1.086603 & 68.483371\\
        529107638353525376 & 1.086814 & 68.481625\\
        529106508781158656 & 1.280217 & 68.513387\\
        \hline
        \multicolumn{3}{c}{SFO 5}\\
        513565148141778560 & 37.017032 & 61.575681\\
        513561716467377536 & 37.180873 & 61.526220\\
        513560926193404544 & 37.181123 & 61.494877\\
        513567209726131968 & 37.211332 & 61.665927\\
        513562609820646656 & 37.256233 & 61.558057\\
        513562919058291584 & 37.309085 & 61.557092\\
        513586318040856064 & 37.423374 & 61.588046\\
        \hline
        \multicolumn{3}{c}{SFO 7}\\
        465931972268206208 &38.472445 &61.947325\\
        465919808920854016 & 38.490779 & 61.887787\\
        465918228372937600 & 38.558287 & 61.801810\\
        465915062974574080 & 38.592096 & 61.757173\\
        465915681452980736 & 38.633960 & 61.812643\\
        465915612730624896 & 38.648419 & 61.810991\\
        465914066542002688 & 38.673421 & 61.717098\\
        465915406574942848 & 38.675025 & 61.782352\\
        465922385902976640 & 38.765262 & 61.914548\\
        \hline
        \multicolumn{3}{c}{SFO 11}\\    
        464554455998190592 & 42.856584 & 60.101325\\
        464554421638442752 & 42.905852 & 60.107391\\
        464555830388653568 & 42.965128 & 60.125149\\
        464554284199714560 & 42.965341 & 60.112129\\
        464555830387938048 & 42.967249 & 60.119478\\
        464555830387929600 & 42.975882 & 60.128772\\
        \hline
        \multicolumn{3}{c}{SFO 17}\\ 
         3340887246596977664 & 82.375789 & 12.098617\\
3340893641804344320 & 82.469851 & 12.163148\\ 
3340894088480934400 & 82.490234 & 12.201947\\ 
3340894500797788672 & 82.502617 & 12.229013\\ 
3340882749767481344 & 82.515892 & 12.022681\\ 
3340892714091577856 & 82.554715 & 12.146044\\ 
3340896184424945664 & 82.570998 & 12.261384\\ 
3340888380468287488 & 82.608051 & 12.055951\\ 
3340889106319041024 & 82.650334 & 12.111273\\ 
3340890480708558848 & 82.715427 & 12.143540\\ 
    \hline
\end{tabular}\\
\end{table}

\begin{table}
\contcaption{}
    \begin{tabular}{ccc}
		\hline
		 candidate YSOs (\textit{Gaia} EDR3) & RA & Dec.\\
		(1)&(2)&(3)\\
\hline 
\multicolumn{3}{c}{SFO 17}\\ 
3340878179922230272 & 82.754191 & 12.070638\\ 
3340878381785847296 & 82.754855 & 12.096842\\ 
3340878351720911872 & 82.783418 & 12.101767\\         
3340878935836457472 & 82.802044 & 12.131820\\ 
3340973111582846464 & 82.814597 & 12.189917\\ 
3340877935108760576 & 82.822064 & 12.096582\\ 
3340972806641584128 & 82.831061 & 12.154255\\ 
3340972733626479616 & 82.835537 & 12.154705\\ 
3340877728950818816 & 82.839460 & 12.095517\\ 
3340877728950193152 & 82.840267 & 12.096578\\ 
3340874224256986112 & 82.840446 & 12.043069\\ 
3340871028801247232 & 82.841301 & 11.915700\\ 
3340871067456399360 & 82.849493 & 11.927072\\ 
3340973558259384832 & 82.853259 & 12.202055\\ 
3340973631275296256 & 82.856891 & 12.214288\\ 
3340973352101788672 & 82.867039 & 12.197495\\ 
3340973596915556864 & 82.868331 & 12.209797\\ 
3340678446763632640 & 82.910861 & 11.807321\\ 
3340681981520082944 & 82.916445 & 11.847166\\ 
3340972497403927552 & 82.918674 & 12.179685\\ 
3340658415036209664 & 82.938609 & 11.631929\\ 
3340975551124339712 & 82.963700 & 12.272423\\ 
3341000736812880896 & 82.964315 & 12.554071\\ 
3340994556355913728 & 82.986524 & 12.419345\\ 
3340993869161153536 & 82.993690 & 12.379817\\ 
3340969061430118400 & 82.999760 & 12.135219\\ 
3341001256504863744 & 83.002221 & 12.570847\\ 
3340997064616797696 & 83.025877 & 12.473332\\ 
3340997133336271360 & 83.026715 & 12.489470\\ 
3340681088166923776 & 83.030248 & 11.901174\\ 
3340779838055977984 & 83.043450 & 12.050434\\ 
3340984072339594240 & 83.060387 & 12.408031\\ 
3340969886063797248 & 83.064392 & 12.159503\\ 
3340984042275960832 & 83.081278 & 12.406744\\ 
3340774069916236800 & 83.109319 & 11.906737\\ 
3340774065621826048 & 83.111576 & 11.911601\\ 
3340984695110967296 & 83.143467 & 12.457907\\ 
3341008300251213824 & 83.169933 & 12.487643\\ 
3340984454592793088 & 83.179992 & 12.451460\\ 
3341007922294096896 & 83.189609 & 12.450624\\ 
3340764311750480384 & 83.266538 & 11.968315\\ 
3340761528611688960 & 83.287080 & 11.898705\\ 
\hline
    \multicolumn{3}{c}{SFO 18}\\ 
    3336168314489240576 & 85.660435 & 9.310332 \\
3336168043908536832 & 85.680581 & 9.287969\\ 
3336155189070384128 & 85.729819 & 9.246989\\ 
3336152375867856384 & 85.747103 & 9.164555\\ 
3336161549917961216 & 85.802402 & 9.283812\\ 
3336180585212965376 & 85.814037 & 9.491866\\ 
3336149519712868864 & 85.837176 & 9.101955\\ 
3336160690924950528 & 85.873411 & 9.283380\\ 
3336177454181082624 & 85.890385 & 9.496435\\ 
3336156121079561216 & 85.914989 & 9.178026\\ 
3336149867606924288 & 85.935460 & 9.133633\\ 
3336156292877976064 & 85.961132 & 9.206919\\ 
3336157422453353472 & 85.963992 & 9.227901\\ 
3336159419614147584 & 85.968111 & 9.283435\\ 
3336157873425925120 & 85.968650 & 9.266585\\ 
3336102897844570112 & 85.972911 & 9.119901\\ 
3336159557053093376 & 85.985790 & 9.304671\\ 
3336159282175185920 & 86.006802 & 9.289093\\ 
3336159071720837120 & 86.008674 & 9.260364\\ 
\hline
\end{tabular}\\
\end{table}

\begin{table}
\contcaption{}
    \begin{tabular}{ccc}
		\hline
		 candidate YSOs (\textit{Gaia} EDR3) & RA & Dec.\\
		(1)&(2)&(3)\\
\hline 
\multicolumn{3}{c}{SFO 18}\\ 
3336107772631258624 & 86.030253 & 9.110574\\
3336109048235617280 & 86.041829 & 9.144624\\ 
3336109563633776640 & 86.044381 & 9.212530\\ 
3336107429033839616 & 86.053240 & 9.086216\\ 
3336108223604259328 & 86.061986 & 9.135564\\ 
3336111006743055360 & 86.079400 & 9.209260\\ 
3336104581471997440 & 86.080668 & 9.089784\\ 
3336158594980405248 & 86.080847 & 9.271995\\ 
3336158521965029888 & 86.087587 & 9.267106\\ 
3336108361043207168 & 86.090606 & 9.148215\\ 
3336108361043207168 & 86.091194 & 9.147915\\ 
3336109937294863872 & 86.095395 & 9.176425\\ 
3336110834944361984 & 86.107235 & 9.200320\\ 
3336105028148588032 & 86.108237 & 9.116840\\ 
3336109902935129088 & 86.112112 & 9.178385\\ 
3336104955132704768 & 86.119424 & 9.105122\\ 
3336256176637551104 & 86.137795 & 9.446547\\ 
3336110525706708992 & 86.154261 & 9.222194\\ 
3336104787630415872 & 86.157234 & 9.099283\\ 
3336256103621173248 & 86.158848 & 9.432716\\ 
3336107227171825664 & 86.194115 & 9.189299\\ 
3336204155993987072 & 86.220057 & 9.218111\\ 
3335326299037613056 & 86.226617 & 8.713875\\ 
3336093513339795456 & 86.267100 & 9.081243\\ 
\hline
    \multicolumn{3}{c}{SFO 25}\\ 
        3326929118281227008 &100.205037 & 9.960728\\
        3326717466589159424 & 100.207263 & 9.882923\\
        3326928705964373888 & 100.210787 & 9.915904\\
        3326740728129988096 & 100.225458 & 9.873459\\ 
        3326928847700236928 & 100.225790 & 9.931090\\        
        3326932004499068928 & 100.249125 & 10.036809\\
        3326943897265235840 & 100.256368 & 10.248875\\
        3326935960166062720 & 100.261885 & 10.119963\\
        3326930874924790656 & 100.280257 & 9.975301\\
        3326943927328338304 & 100.281154 & 10.251207\\
        3326740521973590656 & 100.321865 & 9.908974\\ 
        \hline
    \multicolumn{3}{c}{SFO 26}\\ 
        3046018985610505600 & 105.931990 & -11.615694\\
        3046026540451631232 & 105.957919 & -11.521443\\
        3046025720119222784 & 105.974974 & -11.546624\\
        3046025715819957248 & 105.975102 & -11.543665\\
        3046025548320530304 & 105.991885 & -11.544413\\
        3046025440943675008 & 105.993075 & -11.559610\\
        3045831583295956224 & 106.005495 & -11.604856\\
        3045831617655703168 & 106.016364 & -11.599203\\
        3045830346345248384 & 106.022991 & -11.664542\\
        \hline
        \multicolumn{3}{c}{SFO 27}\\
        3046026540451631104 & 105.957919 & -11.521443\\
        3046030427403041792 & 105.968721 & -11.438005\\ 
        3046026750911057408 & 105.970952 & -11.493117\\ 
        3046025715819957248 & 105.975102 & -11.543665\\ 
        3046027575540896256 & 105.984839 & -11.428186\\ 
        3046030564841979904 & 105.987988 & -11.409098\\ 
        3046027609900606976 & 106.005055 & -11.425314\\ 
        3046032072370037248 & 106.006838 & -11.401754\\ 
        3046027541185004544 & 106.009740 & -11.427599\\ 
        3046027335022717952 & 106.017002 & -11.443177\\ 
        3046032179749668864 & 106.019563 & -11.394363\\ 
        3046029224812170240 & 106.024675 & -11.399645\\ 
        3046032656484579328 & 106.025013 & -11.357863\\   
        3046032175450394624 & 106.025134 & -11.387689\\ 
        \hline
        \end{tabular}\\
\end{table}

\begin{table}
\contcaption{}
    \begin{tabular}{ccc}
		\hline
		 candidate YSOs (\textit{Gaia} EDR3) & RA & Dec.\\
		(1)&(2)&(3)\\
\hline 
\multicolumn{3}{c}{SFO 27}\\ 
        3046032214109395968 & 106.027125 & -11.374434\\ 
        3046032179749663232 & 106.027349 & -11.387875\\ 
        3046033829017076224 & 106.027389 & -11.309653\\ 
        3046028812495321088 & 106.027897 & -11.435688\\ 
        3046032901304147456 & 106.034603 & -11.334795\\ 
        3046033000084199424 & 106.038398 & -11.321701\\ 
        3046029259171897344 & 106.041477 & -11.387892\\ 
        3046408414583416832 & 106.051334 & -11.299733\\ 
        3046029667189515264 & 106.056360 & -11.370099\\ 
        3046028984293995520 & 106.057986 & -11.411538\\ 
        3046029568409539584 & 106.059362 & -11.388029\\ 
        3046026235515287552 & 106.064109 & -11.478023\\ 
        3046028915574513664 & 106.070001 & -11.409016\\ 
        3046029705848479744 & 106.083301 & -11.372899\\ 
        3046029396610840064 & 106.085899 & -11.387360\\ 
        3046027678624293888 & 106.094784 & -11.464144\\ 
        3046028571977483776 & 106.096879 & -11.404809\\ 
        3046028567676658688 & 106.107098 & -11.403421\\ 
        3045840688626813952 & 106.169933 & -11.426916\\ 
        3045840310669678080 & 106.187495 & -11.466389\\
        \hline
        \multicolumn{3}{c}{SFO 31}\\ 
        2163139941968363520&312.655751 & 44.348055\\
        2163139873248885632&312.668531 & 44.339344\\
        2163140006394622208&312.677957 & 44.364693\\
        2163139804527939968&312.702434 & 44.351252\\
        2163139804529410944&312.702892 & 44.348122\\
        2163140457364449920&312.723076 & 44.407410\\
        2163140560443665152&312.723456 & 44.408447\\
        2163140143827638272&312.724408 & 44.361061\\
        2163136810933010048&312.728592 & 44.336789\\
        2163140251206016640&312.728724 & 44.382835\\
        2163136712151471360&312.734157 & 44.306060\\
        2163140281266601728&312.744801 & 44.388165\\
        2163136746512695424&312.745882 & 44.331735\\
        2163136746511218176&312.747839 & 44.327502\\
        2163137120170654720&312.758957 & 44.335850\\
        2163140319925497088&312.763891 & 44.404285\\
        2163136192456870016&312.831063 & 44.325146\\
        \hline
        \multicolumn{3}{c}{SFO 36}\\ 
        2178434217440720384 & 323.852133 & 57.550292\\
2178421813575206400 & 323.905550 & 57.478271\\ 
2178434389239575040 & 323.938735 & 57.557456\\ 
2178417656046880256 & 323.957301 & 57.401123\\ 
2178434561038259712 & 323.960316 & 57.596456\\ 
2178415834982840832 & 323.985635 & 57.347983\\ 
2178446002831454720 & 324.067038 & 57.580120\\ 
2178418167132388864 & 324.070784 & 57.444386\\ 
2178446342117152768 & 324.077924 & 57.607406\\ 
2178444864649304576 & 324.093245 & 57.528092\\ 
2178445109478268416 & 324.098667 & 57.545841\\ 
2178418304571527680 & 324.104485 & 57.463934\\ 
2178444899007898368 & 324.118437 & 57.537037\\ 
2178417278089774080 & 324.145473 & 57.436727\\ 
2178445624874345472 & 324.153753 & 57.568306\\ 
2178440741484235264 & 324.158434 & 57.449372\\ 
2178441604784967168 & 324.160031 & 57.488165\\ 
2178441909720014336 & 324.163107 & 57.498105\\ 
2178441914022611456 & 324.170965 & 57.502272\\ 
2178441707864184832 & 324.191461 & 57.492777\\ 
2178441742223922688 & 324.198437 & 57.498325\\ 
2178440470906711552 & 324.204093 & 57.421181\\ 
2178369930368001536 & 324.206879 & 57.374168\\ 
\hline
        \end{tabular}\\
\end{table}

\begin{table}
\contcaption{}
    \begin{tabular}{ccc}
		\hline
		 candidate YSOs (\textit{Gaia} EDR3) & RA & Dec.\\
		(1)&(2)&(3)\\
\hline 
\multicolumn{3}{c}{SFO 36}\\ 
2178442115878750464 & 324.211327 & 57.519617\\ 
2178440913279379456 & 324.224007 & 57.478980\\ 
2178440814510995456 & 324.224779 & 57.466286\\ 
2178440642705389824 & 324.239399 & 57.458397\\ 
2178393398069567488 & 324.239518 & 57.389290\\ 
2178440642712309760 & 324.240275 & 57.459183\\ 
2178442081510495232 & 324.240962 & 57.518048\\ 
2178441291236505088 & 324.241322 & 57.486287\\ 
2178442154540783104 & 324.247783 & 57.526332\\ 
2178393604228050432 & 324.253665 & 57.422881\\ 
2178442257613053952 & 324.254910 & 57.534543\\ 
2178442459476372480 & 324.255522 & 57.558474\\ 
2178441226827857920 & 324.256374 & 57.479491\\ 
2178440573992845568 & 324.260620 & 57.437321\\ 
2178441226827857920 & 324.262816 & 57.483628\\ 
2178440573992840448 & 324.263380 & 57.455095\\ 
2178447102343234048 & 324.265886 & 57.624686\\ 
2178393634282935040 & 324.274296 & 57.425248\\ 
2178442188900684032 & 324.275295 & 57.533776\\ 
2178440951949966848 & 324.279290 & 57.450210\\ 
2178442184589714944 & 324.282137 & 57.536404\\ 
2178444078686267648 & 324.283960 & 57.604479\\ 
2178441364259874304 & 324.284523 & 57.506525\\ 
2178393569868318208 & 324.287895 & 57.430134\\ 
2178441261187597056 & 324.289007 & 57.496765\\ 
2178441261187598336 & 324.289150 & 57.494766\\ 
2178441432986447360 & 324.293050 & 57.522256\\ 
2178441432986447872 & 324.293934 & 57.520104\\ 
2178394016544915456 & 324.294569 & 57.436194\\ 
2178441055029170688 & 324.297757 & 57.491271\\ 
2178443769448645376 & 324.298756 & 57.558096\\ 
2178394016544908288 & 324.300637 & 57.457276\\ 
2178443769448645632 & 324.301422 & 57.558842\\ 
2178394050904648448 & 324.307880 & 57.457498\\ 
2178440982007912704 & 324.311440 & 57.470808\\ 
2178441158101474560 & 324.312343 & 57.486782\\ 
2178441158108390144 & 324.312770 & 57.486519\\ 
2178394050904653312 & 324.316308 & 57.449764\\ 
2178393982185180160 & 324.318140 & 57.444534\\ 
2178444113039290368 & 324.319114 & 57.605811\\ 
2178393947825444864 & 324.320866 & 57.438297\\ 
2178441085078071808 & 324.321454 & 57.479803\\ 
2178441089388915200 & 324.322358 & 57.489084\\ 
2178441089388913920 & 324.322583 & 57.490908\\ 
2178441192468289024 & 324.331546 & 57.513574\\ 
2178441089388920832 & 324.333159 & 57.479476\\ 
2178442876095464192 & 324.333704 & 57.526427\\ 
2178392229838569472 & 324.336028 & 57.350092\\ 
2178442669938739712 & 324.341791 & 57.512066\\ 
2178393982185185792 & 324.342839 & 57.445544\\ 
2178442876095465472 & 324.348673 & 57.531615\\ 
2178442974872913152 & 324.349750 & 57.548867\\ 
2178392917033317376 & 324.350403 & 57.403143\\ 
2178394188343600896 & 324.353173 & 57.485799\\ 
2178442704296780288 & 324.363834 & 57.524861\\ 
2178442699985787904 & 324.364255 & 57.523214\\ 
2178392848313851392 & 324.368435 & 57.390246\\ 
2178443941247336192 & 324.370589 & 57.601205\\ 
2178443013534417408 & 324.371514 & 57.548204\\ 
2178395661507500800 & 324.374572 & 57.487368\\
2178443112301672704 & 324.377703 & 57.583804\\ 
2178443116606937088 & 324.378112 & 57.574356\\ 
2178393054472271872 & 324.378334 & 57.416702\\ 
\hline
        \end{tabular}\\
\end{table}

\begin{table}
\contcaption{}
    \begin{tabular}{ccc}
		\hline
		 candidate YSOs (\textit{Gaia} EDR3) & RA & Dec.\\
		(1)&(2)&(3)\\
\hline 
\multicolumn{3}{c}{SFO 36}\\ 
2178393054472273920 & 324.384810 & 57.417549\\ 
2178395597093125120 & 324.384830 & 57.490160\\ 
2178395592793859072 & 324.386841 & 57.474184\\ 
2178444250484986880 & 324.389828 & 57.597511\\ 
2178442807375996160 & 324.390869 & 57.537961\\ 
2178442837424749056 & 324.398815 & 57.549550\\ 
2178395425294441472 & 324.407814 & 57.479645\\ 
2178442738656526336 & 324.410436 & 57.527991\\ 
2178395631452868096 & 324.411854 & 57.493597\\ 
2178442841735735296 & 324.411907 & 57.547055\\ 
2178443425844596224 & 324.418660 & 57.575885\\ 
2178443219692854528 & 324.420668 & 57.560143\\ 
2178395356574972928 & 324.420927 & 57.466933\\ 
2178450637088253696 & 324.424351 & 57.677820\\ 
2178443219692855040 & 324.424390 & 57.560675\\ 
2178443421539290880 & 324.425704 & 57.575414\\ 
2178443219693138688 & 324.428171 & 57.556967\\ 
2178443425852969728 & 324.428688 & 57.579511\\ 
2178395837611290112 & 324.435646 & 57.528416\\ 
2178392740924572672 & 324.437010 & 57.403748\\ 
2178389339310069760 & 324.438109 & 57.328373\\ 
2178395356574976256 & 324.438442 & 57.471469\\ 
2178443254045917696 & 324.442181 & 57.574461\\ 
2178395837599325696 & 324.445379 & 57.532251\\ 
2178394497581292800 & 324.446132 & 57.434589\\ 
2178395459654186240 & 324.452937 & 57.489681\\ 
2178389790297157120 & 324.453941 & 57.389113\\ 
2178396142543875840 & 324.455299 & 57.532880\\ 
2178443185333123584 & 324.459138 & 57.561200\\ 
2178395562733396992 & 324.459288 & 57.502631\\ 
2178394291422867968 & 324.459296 & 57.430166\\ 
2178394699429915904 & 324.462830 & 57.463914\\ 
2178394531941272064 & 324.465701 & 57.447897\\ 
2178443284100355072 & 324.486562 & 57.580033\\ 
2178391302125646848 & 324.489868 & 57.405441\\ 
2178391199046443520 & 324.490090 & 57.379878\\ 
2178395975050256896 & 324.492197 & 57.522179\\ 
2178394669380222464 & 324.502463 & 57.473688\\ 
2178450538317241344 & 324.514567 & 57.693017\\ 
2178390958528283648 & 324.524777 & 57.378809\\ 
2178402262882609920 & 324.532163 & 57.598123\\ 
2178396009413394176 & 324.534021 & 57.524052\\ 
2178394394502336768 & 324.535220 & 57.446542\\ 
2178391370845370880 & 324.535333 & 57.419930\\ 
2178449232647211008 & 324.535644 & 57.618771\\ 
2178395081697083392 & 324.540760 & 57.495191\\ 
2178391130327213568 & 324.541678 & 57.397955\\ 
2178394768154306560 & 324.542438 & 57.452310\\ 
2178395253495768320 & 324.547566 & 57.517118\\ 
2178391370845377024 & 324.550076 & 57.416853\\ 
2178395287855504384 & 324.557715 & 57.528134\\ 
2178390305693267456 & 324.562898 & 57.365394\\ 
2178449816762756864 & 324.571359 & 57.657436\\ 
2178390305693509376 & 324.572937 & 57.375203\\ 
2178391881932722688 & 324.599781 & 57.460037\\ 
2178390649290896384 & 324.605400 & 57.386767\\ 
2178402056712608768 & 324.608188 & 57.569242\\ 
2178391989320655360 & 324.609953 & 57.477909\\ 
2178391680083024640 & 324.612183 & 57.444025\\ 
2178400888493119488 & 324.614296 & 57.518909\\ 
2178400888493120512 & 324.618131 & 57.518649\\ 
2178401232090491904 & 324.626306 & 57.548643\\ 
2178391611363553280 & 324.626488 & 57.438370\\ 
\hline
        \end{tabular}\\
\end{table}

\begin{table}
\contcaption{}
    \begin{tabular}{ccc}
		\hline
		 candidate YSOs (\textit{Gaia} EDR3) & RA & Dec.\\
		(1)&(2)&(3)\\
\hline 
\multicolumn{3}{c}{SFO 36}\\ 
2178391607051821568 & 324.634026 & 57.443286\\ 
2178397864836157440 & 324.635643 & 57.504446\\ 
2178391611363558144 & 324.640927 & 57.434762\\ 
2178401025932067328 & 324.645041 & 57.547196\\ 
2178397899195898368 & 324.646700 & 57.503741\\ 
2178398002263548416 & 324.655419 & 57.517729\\ 
2178397830476432896 & 324.664847 & 57.487861\\ 
2178397521238551040 & 324.665652 & 57.464661\\ 
2178401640101582848 & 324.667773 & 57.573126\\ 
2178397516931971840 & 324.681268 & 57.457493\\ 
2178397551292322304 & 324.689294 & 57.473025\\ 
\hline 
\multicolumn{3}{c}{SFO 37}\\ 
        2178110961019859200 & 325.111487 & 56.606547\\
        2178110961019859328 & 325.113856 & 56.606019\\
        2178110961019860864 & 325.119532 & 56.602676\\
\hline
\multicolumn{3}{c}{SFO 38}\\ 
        2178572790254904192&324.985110 & 58.229860\\
        2178572515384121344&324.993569 & 58.204090\\
        2178573198273201920&325.018974 & 58.253206\\
        2178561309803684608&325.090679 & 58.245960\\
        2178561309803669376&325.106798 & 58.241177\\
        2178561245389942016&325.113851 & 58.239224\\
        2178562791578166784&325.116659 & 58.253957\\ 
        2178574744461539200&325.131643 & 58.298669\\
        2178562516700266624&325.152326 & 58.229408\\
        2178562551060090624&325.153795 & 58.243853\\
        2178562752912981248&325.154343 & 58.250826\\        
        2178562447980793088&325.161832 & 58.217437\\
        2178562482340616192&325.171263 & 58.233038\\
        2178562654139305856&325.171512 & 58.253125\\        
        2178575156779164416&325.178405 & 58.327041\\
        2178562649833718528&325.186884 & 58.250933\\
        2178561412882638336&325.187700 & 58.187668\\
        2178562654140703872&325.193728 & 58.256421\\
        2178562684193256704&325.200132 & 58.260469\\
        \hline
        \multicolumn{3}{c}{SFO 39}\\ 
        2178252591856861440&326.315947 & 57.398345\\
        2178253347771102336&326.392006 & 57.449805\\
        2178253343467691904&326.403087 & 57.454570\\
        2178302310399899136&326.460928 & 57.569224\\
        2178254619081525248&326.475266 & 57.471767\\
        2178249052803834240&326.501131 & 57.385957\\
        2178254722160739328&326.506666 & 57.493876\\
        2178255336331599104&326.567890 & 57.537926\\
        2178254065020940544&326.608237 & 57.474642\\
        \hline
        \multicolumn{3}{c}{SFO 54}\\ 
        5528463903210289152&128.825305 & -40.639072\\
        5528463898908631296&128.827379 & -40.638469\\
        5528460982632433536&128.851861 & -40.630607\\        
        5528462460101182336&128.857268 & -40.625275\\
        5528460913912957952&128.857845 & -40.649441\\
        5528460909612659840&128.858711 & -40.647091\\
        5528462391381705344&128.865797 & -40.629472\\
        5528460600376692224&128.878977 & -40.683239\\
        5528462215282558848&128.886545 & -40.637564\\
        5528459814401329152&128.900026 & -40.685088\\
        5528461326229815296&128.917863 & -40.668665\\
        5528461493728727808&128.919382 & -40.643420\\
        \hline
        \end{tabular}\\
\end{table}

\begin{table}
\contcaption{}
    \begin{tabular}{ccc}
		\hline
		 candidate YSOs (\textit{Gaia} EDR3) & RA & Dec.\\
		(1)&(2)&(3)\\
\hline 
        \multicolumn{3}{c}{SFO 55}\\ 
        5525309679224303616&130.258108 & -40.851611\\
        5525309404343867008&130.272932 & -40.879715\\
        5525309404347345664&130.278144 & -40.871491\\
        5525315417300607744&130.297572 & -40.834450\\
        5525315348582080640&130.306144 & -40.835407\\
        5525314558307488384&130.312112 & -40.864011\\
        5525314592666887040&130.319382 & -40.858090\\
        5525315451660343296&130.321984 & -40.833414\\
        5525314730105835904&130.342176 & -40.840931\\
        \hline
        \multicolumn{3}{c}{SFO 64}\\ 
        5339406344853783168&168.040164 & -58.777692\\
        5339406344853784064&168.043617 & -58.775652\\
        5339406276134319872&168.083076 & -58.768528\\
        5339407792307371264&168.083896 & -58.735824\\
        5339406138695368448&168.090774 & -58.789143\\
        5339406276134323840&168.095506 & -58.768474\\
        5339385316693599232&167.970779 & -58.830521\\
        5339406173055110912&168.100420 & -58.779458\\
        5339407719243340544&168.111612 & -58.750773\\
        5339406211759383040&168.125793 & -58.779075\\
        \hline
        \multicolumn{3}{c}{SFO 65}\\ 
        5333619447671205248 &173.115031 & -63.463595\\
        5333606592810294656 & 173.268299 & -63.514073\\
        5333612537045149440 & 173.276352 & -63.482744\\
        \hline
        \multicolumn{3}{c}{SFO 75}\\ 
        5884681167820754944 &238.922042 & -54.662066\\
        5884681270898481408 & 238.933403 & -54.649745\\
        5884681618820301184 & 238.974183 & -54.635937\\
        5884677972363578368 & 239.004325 & -54.691620\\
        \hline
        \multicolumn{3}{c}{SFO 79}\\ 
        5940872985209320960 &249.965948 & -48.861551\\
        5940872985201536640 & 249.973811 & -48.862893\\
        5940862986517159552 & 250.041070 & -48.794521\\
        5940860130322500992 & 250.065560 & -48.913008\\
        5940863192675559808 & 250.071388 & -48.789952\\
        5940862677279451648 & 250.079483 & -48.836526\\
        5940861783926211200 & 250.089274 & -48.878902\\ 
        5940860203378210816 & 250.095269 & -48.908659\\
        5940860199040422912 & 250.100495 & -48.912007\\
        5940861955724931328 & 250.112240 & -48.839791\\
        \hline
        \multicolumn{3}{c}{SFO 82}\\ 
        5968283535216265088 &251.786509 &  -41.270942\\
        5968283397777300352 & 251.818542 & -41.275009\\
        5968283393435690752 & 251.828189 & -41.269569\\
        5968283874470676352 & 251.845438 & -41.231744\\
        5968271681105309312 & 251.850580 & -41.272716\\
        5968271990342949760 & 251.865275 & -41.274959\\
        \hline
        \multicolumn{3}{c}{SFO 89}\\ 
        4066264896044933888 &272.460228 & -24.089038\\
        4066264582463138048 & 272.463524 & -24.097850\\
        4066261528767982336 & 272.474727 & -24.125813\\
        4066264616822113536 & 272.482763 & -24.102785\\
        4066261666237504640 & 272.484526 & -24.113703\\
        4066264616822117248 & 272.485582 & -24.108377\\
        4066264621175030912 & 272.489631 & -24.102080\\
        4066261563127704064 & 272.490897 & -24.121382\\
        4066261661884621952 & 272.491424 & -24.115998\\
        4066261696237927808 & 272.501261 & -24.108107\\
        \hline
	\end{tabular}\\
	
\end{table}



\bsp	
\label{lastpage}
\end{document}